\documentstyle[preprint,aps,eqsecnum]{revtex}
\preprint{\vbox{\baselineskip=12pt
\rightline{CGPG-96/5-5}
\rightline{gr-qc/9608017}}}

\def\be{\nopagebreak[3]\begin{equation}}
\def\ee{\end{equation}}
\def\ba{\nopagebreak[3]\begin{eqnarray}}
\def\ea{\end{eqnarray}}

\def\nl{\nonumber \\}

\def\r{\rangle}
\def\l{\langle}

\def\Tr{{\rm T}}

\def\d{{\rm d}}
\def\e{{\rm e}}

\newcommand{\T}{\textstyle}

\begin{document}
\title{Photon inner-product and the Gauss linking number}
\author{Abhay Ashtekar${}^{1,2}$\thanks{Electronic address: 
ashtekar@phys.psu.edu}
 and Alejandro Corichi${}^1$\thanks{Electronic address: 
corichi@phys.psu.edu}
}
\address{$1$ Center for Gravitational Physics and Geometry\\
Department of Physics, Penn State,\\ 
University Park, PA 16802, USA
} 
\address{$2$ Erwin Schr\"odinger International Institute 
for Mathematical Sciences\\
Boltzmanngasse 9, A-1090 Vienna, Austria
}
\maketitle

\begin{abstract}

It is shown that there is an interesting interplay between
self-duality, loop representation and knot invariants in the quantum
theory of Maxwell fields in Minkowski space-time. Specifically, in the
loop representation based on self-dual connections, the measure that
dictates the inner-product can be expressed in terms of the Gauss
linking number of thickened loops.

\end{abstract} 
\pacs{0350De, 0370+k, 1115-q}

\section{Introduction}
\label{s1}

Source-free electrodynamics is the simplest physical theory based on
connections. {}From a geometric point of view, a natural set of
observables of this theory is given by holonomies of the connection
around closed loops. It is then natural to ask if topological
invariants associated with loops play a physically significant role in
this description.  For the observable algebra, the answer is in the
affirmative: the fundamental Heisenberg uncertainty relations can be
formulated in terms of the Gauss-linking number (see, e.g.,
\cite{ar1,ac} and references contained therein). One can ask if such 
a topological invariant of loops also plays a role in the description
of the Hilbert space of quantum states. The purpose of this note is to
show that the answer is again in the affirmative. Furthermore, the
specific calculation we wish to present is based on an interesting
interplay between self-duality, loop representation and knot theory
and may well be a reflection of a deeper structure that underlies
these three notions.

The basic idea is the following. The standard Fock description of
photons can be reformulated in terms of loops, so that the states can
be regarded as functionals of loops (rather than connections) (see,
e.g., \cite{ar2} and chapter 14 in \cite{aa:91}). There are, however,
several such loop representations. In the one most directly related to
the Fock-Bargmann representation \cite{ar2}, it is the negative
frequency electric field that is diagonal and the conjugate operator
represents the holonomy of positive frequency connections.
Alternatively, one can work with real electric fields and
connections. But then to obtain the Fock representation, the loops
have to be thickened.  In this paper, we will work with yet another
choice: our loop representation will be based on {\it real} electric
fields but {\it self-dual} connections (without any reference to
positive and negative frequencies). Again, to get the Fock
representation, we will have to thicken our loops.  Thus, quantum
states are expressed as functionals of thickened loops and the basic
operators are the holonomies of self-dual connections and real
electric fields. The measure that dictates the inner-product in this
representation has a Gaussian form where the exponent is given by the
self-linking number of thickened loops.

More precisely, the situation is the following. Given a loop $\alpha$
and a weighting function $f$ on $R^3$, we can define a ``canonical
thickening'' $\alpha_f$ of the loop. The self-linking number of a
thickened loop can be computed by adding the Gauss linking numbers of
the loops involved in the thickening with weights given by $f$. The
measure that dictates the inner product is just the exponential of
this self-linking number of $\alpha_f$.  Now, as is well-known, the
Fock inner product depends on the Minkowski metric.  It is quite
interesting that one can put all the information about the space-time
metric in the construction that associates quantum states with
functionals of loops and then express the inner-product itself in
terms of the Gauss linking number which is a topological invariant. A
further striking fact is that this ``coding'' of the inner-product
information in a topological invariant works only in the loop
representation based on {\it self-dual} (or anti-self-dual)
connections.

The plan of the paper is as follows. In section \ref{s2}, we will
collect some mathematical preliminaries. These are used in section
\ref{s3} to construct the loop representation based on self-dual
connections.  The main result then follows in section \ref{s4}. Since
our primary goal is to bring out the interplay between self-duality of
connections, loop representations and the linking number, we will keep
the functional analytic details to a minimum. However, it should be
rather straightforward to see how one can complete our discussion to
obtain a rigorous treatment. Throughout the article, we use units
where $c=1$, but write $\hbar$ and $e$ explicitly.

\section{Mathematical Preliminaries}
\label{s2}
This section is divided in to two parts. In the first, we recall the
phase space formulation of the Maxwell field using self-dual variables
and in the second we introduce the notion of ``form factors''
associated with loops and thickened loops.

\subsection{Self-Dual Variables for the Maxwell Field}
\label{s2.1}
Let us begin with a brief summary of the standard phase space
formulation of Maxwell fields. Denote by $\Sigma$ a space-like
three-plane in Minkowski space, and by $q_{ab}$, the induced positive
definite (flat) metric thereon.  The configuration variable for the
Maxwell field is generally taken to be the connection one-form
$A_a(x)$ (the vector potential for the magnetic field) on $\Sigma$.
Its canonically conjugate momentum is the electric field ${E}^a(x)$ on
$\Sigma$. ($E^a(x)$ naturally arises as a vector density.  However,
since we have an underlying metric, $q_{ab}$, which can be used to add
or remove density weights, we will ignore density weights in this
paper.) The fundamental Poisson bracket is:
\be \{A_a(x), {E}^b(y)\}=\delta^b_a \delta^3(x,y)\ee
The system has one first class constraint, $\partial_a
{E}^a(x)=0$. One can therefore pass to the reduced phase space by
fixing transverse gauge. The true degrees of freedom are then
contained in the pair $(A_a^\Tr (x), {E}^a_\Tr (x))$ of transverse
(i.e.  divergence-free) vector fields on $\Sigma$. Denote by $\Gamma$
the phase space spanned by these fields. On $\Gamma$, the only
non-vanishing fundamental Poisson bracket is:
\be \{A^\Tr_a(x), {E}^b_\Tr(y)\}=\delta^b_a \delta^3(x,y) - 
\Delta^{-1}\partial^b\partial_a  \delta^3(x,y)\, ,
\label{poisbra}\ee
where $\Delta$ is the Laplacian operator compatible with the flat
metric $q_{ab}$.  It is convenient to write $A_a^\Tr (x)$ and
${E}^a_\Tr (x)$ in terms of their Fourier decomposition. Then, the
true degrees of freedom are contained in the new dynamical variables
$q_j(k)$, $p_j(k)$ with $j=1,2$:
\ba A^\Tr_a(x) & = & \frac{1}{(2\pi )^{3/2}} \int \d^3k\,\e^{ik\cdot x}
(q_1(k)m_a(k) + q_2(k)\bar{m}_a(k))\\
{E}^a_\Tr (x) & = & \frac{1}{(2\pi )^{3/2}} 
\int \d^3k\,\e^{ik\cdot x}(p_1(k)m^a(k) + p_2(k)\bar{m}^a(k))\ea
where $m_a$ and $\bar{m}_a$ are transverse (complex) vectors
satisfying: $m_ak^a=0$, and $m_a\bar{m}^a=1$.  The Poisson brackets
(\ref{poisbra}) for the transverse components are,
\be \{q_i(-k), p_j(k^\prime)\} = - \delta_{ij} 
\delta^3(k,k^\prime)\, , \ee
while the fact that $A^\Tr_A(x)$ and ${E}^a_\Tr(x)$ are real
translates to the ``reality conditions'':
\be \bar{q_i}(k)  =  q_i(-k)\, \quad {\rm and} \quad 
\bar{p_i}(k)  = p_i(-k) \label{reality} \, .\ee

In order to construct the self dual connection, we will use
$d^\Tr_a(x)$, the transverse vector potential of the electric field
$({E}^a_\Tr (x)={\epsilon}^{abc}\partial_bd^\Tr_c(x))$
\be d^\Tr_a(x)= \frac{1}{(2\pi )^{3/2}} \int \frac{\d^3k}{|k|}\, 
\e^{ik\cdot x}(p_1(k)m_a(k)- p_2(k)\bar{m}_a(k))\ee
Let us define the {\it self dual connection} as
\be {}^\uparrow\!A^\Tr_a(x) := -iA^\Tr_a(x) + d^\Tr_a(x)\ee
We want to use the pair (${}^\uparrow\!A^\Tr_a(x)$, ${E}^a_\Tr(x)$) as
the basic variables.  In terms of the $(q_j, p_j)$ coordinates, the
self dual connection takes the form,
\ba
{}^\uparrow\!A^\Tr_a(x) &=& -\frac{1}{(2\pi )^{3/2}} 
\int \frac{\d^3k}{|k|}\,\e^{ik\cdot x} [(-p_1(k)+i|k|q_1(k))m_a(k) 
+ (p_2(k)+i|k|q_2(k))\bar{m}_a(k)]\nonumber\\
&=& - \frac{1}{(2\pi)^{3/2}} \int \frac{\d^3k}{|k|}\,
\e^{ik\cdot x}[z_1(k)m_a(k) + z_2(k)\bar{m}_a(k)]
\ea
where
\be z_1(k) := -p_1(k) + i |k| q_1(k)\quad {\rm and} \quad
z_2(k) := p_2(k) + i |k| q_2(k)\ee
The basic Poisson brackets for the pairs $(z_i(k), p_j(k))$,  -- the
Fourier components of the self-dual connection and the real electric 
field--  are:
\be \{p_i(k), z_j(-k^\prime)\}= i|k|\delta_{ij}
\delta^3(k,k^\prime)\, ,\ee
and the ``reality conditions'' (\ref{reality}) now become:
\be z_1(-k) + \bar{z}_1(k)  = - 2 p_1(-k)\quad {\rm and} \quad
z_2(-k) + \bar{z}_2(k)  =  2 p_2(-k) \ee
Finally, for later convenience, let us examine the self-dual magnetic
field $B^a := \epsilon^{abc} \partial_b {}^\uparrow\!A_c$.  Its
Fourier components of the magnetic field are given by
\be B_1(k)  =  - z_1(k)\, , \quad {\rm and}\quad 
B_2(k) = z_2(k) \ee
The reality conditions for the magnetic field read
\be B_{j}(k) + \overline{B_{j}(-k)} =  2 p_j(k)
\label{mgreal}\ee
These conditions will play an important role in the selection of the
inner-product in the loop representation.

\subsection{Loops}
\label{s2.2}
Let us begin with some definitions. By a {\it loop} we shall mean a
continuous and piecewise smooth mapping $\gamma$ from $S^1$ to
$\Sigma$, where $s \in [0,2\pi ]$.  Two loops $\gamma$ and $\beta$
will be said to be {\it holonomically equivalent} if, for every smooth
connection $A_a$, we have $\oint_\gamma\, A_a \d s^a = \oint_\beta\,
A_a\d s^a$. It turns out that two holonomically equivalent loops,
$\gamma$ and $\beta$, can differ from each other only through: i)
reparametrization, $\gamma(s)=\beta(s^\prime)$ for some
(orientation-preserving) reparametrization $s\rightarrow s^\prime$ of
the curve $\beta(s)$; ii) retracing identity, $\gamma = l
\cdot \beta \cdot l^{-1}$, where $l$ is a line segment and $\cdot$
indicates composition of segments \cite{al2}.  Each equivalence class
will be referred to as a {\it holonomic loop}.  Since loops will
primarily enter our discussions through holonomies, it is these
equivalence classes --rather than individual loops-- that will be
directly relevant to our discussion.  To keep the notation simple, we
will use the same symbols --say $\gamma$-- to denote both an individual
loop and the holonomic loop it defines; the context should suffice to
resolve the resulting ambiguity.

An analytic characterization of holonomic loops can be given through
certain distributional vector densities, called {\it form factors}.
Given a loop $\alpha$, its {\it form factor}, $F^a(\alpha,x)$, is
defined via:
\be \int \d^3x F^a(\alpha,x)w_a(x) = \oint_{\alpha} w_a \d s^a\ee
Thus, $F^a(\alpha,x)$ may be more directly expressed as
\be F^a(\alpha,x) = \oint_{\alpha}\d s\, \dot{\alpha}^a(s)
\delta^3(x,\alpha(s))\ee
where $\alpha (s)$ is a point on the loop $\alpha$ at parameter value
$s$ and $\dot{\alpha}^a(s)$ the tangent vector to $\alpha$ at
$\alpha(s)$.  Note that the form factor $F^a(\alpha,x)$ is
automatically divergence free,
\be \partial_a F^a(\alpha,x)= 0 \, ,\ee
because $\oint_\alpha \partial_a f\,\d s^a = 0$.  It is often
convenient to perform a Fourier transform to obtain the momentum space
representation of $F^a(\alpha,x)$. We have:
\ba F^a(\alpha,k) & := & \frac{1}{(2\pi )^{3/2}} \int \d^3x\, 
\e^{-i k\cdot x} F^a(\alpha,x)\nl
& = & \frac{1}{(2\pi )^{3/2}} \oint_{\alpha} \d s\, \dot{\alpha}^a(s)
\e^{-ik\cdot \alpha(s)} \ea

Let us note a few properties of these form factors. First, two loops
$\alpha$ and $\beta$ will have the same form factors if and only if
they are holonomic. Thus, $F^a(\alpha,x)$ can be used to characterize
holonomic loop $\alpha$. Next, since $F^a(\alpha,x)$ is
divergence-free its Fourier transform is transverse
$(k_aF^a(\alpha,k)=0)$. We can write the two independent components
as:
\ba F_1(\alpha, k)\equiv F^{+}(\alpha, k) & = & \frac{1}{(2\pi )^{3/2}} 
\oint_{\alpha} \d s\, \dot{\alpha}^a(s) \bar{m}_a(k) 
\e^{-i k\cdot\alpha(s)}
\nonumber\\
F_2(\alpha, k) \equiv F^{-}(\alpha, k) & = & \frac{1}{(2\pi )^{3/2}}
\oint_{\alpha} \d s\,
\dot{\alpha}^a(s)m_a(k) \e^{-i k\cdot \alpha(s)}\ea
so that
\be F^a(\alpha,k) = F^{+}(\alpha ,k)m^a(k) + 
F^{-}(\alpha,k) \bar{m}^a(k) \ee
(We have introduced the $\pm$ notation because in the quantum theory,
$F_1$ will capture positive helicity and $F_2$ the negative.)  This
transversality of form factors will play an important role in the
loop-representation because it captures in a natural way the gauge
invariance of the theory, i.e. the transversality of the photon. The
next property follows from the fact that $F^a(\alpha,x)$ is
real. Consequently, its Fourier transform $F^a(\alpha,k)$ satisfies
the ``reality condition''
\be \overline{F}_j(\alpha, k)= - F_j(\alpha, -k)\label{formreal} \ee
Finally, given two holonomic loops $\alpha$ and $\beta$, we define a
new holonomic loop, $\alpha \# \beta$ as follows: $\alpha \# \beta = l
\cdot \alpha \cdot l^{-1}\cdot \beta$ where $l$ is any line segment 
joining a point on $\alpha$ to a point on $\beta$.  (Because of its
geometric picture $\alpha\#\beta$ is sometimes called the ``eye-glass
loop''.) Using the definition of the form factors, we now have  
\be F_j(\alpha \# \beta), k)= F_j (\alpha, k) + F_j(\beta, k) \ee

In order to construct the quantum theory in the loop representation,
we will need to thicken the loops appropriately. We will conclude this
section by indicating how this can be done. Fix an averaging function
$f_r(x)$ such that $\int_{R^3}d^3x f_r(x)=1$, and goes to a delta
function when $r\rightarrow 0$.  A convenient choice is:
\be f_r(y)= \frac{1}{(2\pi r)^{3/2}}\exp 
\left( -\frac{y^2}{2r}\right) \ee
To make the discussion concrete, we will make this choice and thus
characterize the thickening completely by a real parameter $r$.
(However, the generalization to arbitrary smearing functions is
obvious.)  Now, given a loop $\alpha$ we take the loop $\alpha + y$
obtained by rigidly shifting the loop by the vector $y^a$,
\be (\alpha + y)^a(s) = \alpha ^a(s) + y^a \ee
Next, we can average over $y$ using the weight $f_r(y)$ and define a
``smeared form factor'' via:
\ba F^a_r(\alpha, x) & := & \int \d^3y f_r(y) F^a(\alpha + y, x)
\nonumber\\
& = & \int \d^3y f_r(y) \oint_{\alpha} ds\, \dot{\alpha}^a(s)
\delta^3(x - \alpha (s) - y)
\nonumber\\
& = & \oint_{\alpha} \d s\, \dot{\alpha}^a(s) f_r(x - \alpha(s)) \ea
Its Fourier transform $F^a_r(\alpha, k)$ satisfies
\be F^a_r(\alpha, k)= \exp\left(-\frac{r^2k^2}{2}\right) 
F^a(\alpha, k)\ee
We will see that these $F^a_r(\alpha, k)$ can be used as ``generalized
coordinates'' for loops. More precisely, once the weight functions
$f_r(y)$ are chosen, we can associate with any loop a transverse, {\it
smooth} vector field,
\be\alpha \longrightarrow F^a (k) := F^a_r(\alpha, k)\, .\ee
As is well-known, photon states can be expressed as suitable
functionals, $\Phi[F]$, of smooth vector fields $F^a(k)$ (in the
representation in which the electric field is diagonal). These can be
pulled-back to the loop space to yield functionals $\Psi_r(\alpha)=
\Phi[F]|_{F=F_r(\alpha,k)}$.  Thus, the entire Fock space of photon 
states can be expressed in terms of suitable functionals of loops. 
This fact will be exploited in the next section.

\section{Quantum Theory}
\label{s3}
This section is divided in to three parts. In the first, we recall a
general quantization program (for details, see \cite{aa:rst},
\cite{almmt}), in the second we construct a $\star$-algebra of operators 
based on loop variables and in the third we construct the loop
representation.

\subsection{Quantization Program} 
\label{s3.1}
Consider a classical system with phase space $\Gamma$. To construct
the quantum theory, we can proceed in the following steps.

i) Choose a subspace ${\cal S}$ of the space of complex valued
functions on $\Gamma$ which is closed under the Poisson bracket
operation and large enough so that any well behaved function on
$\Gamma$ can be expressed as (possible the limit of) a sum of products
of elements of ${\cal S}$. Elements of ${\cal S}$ are called {\it
elementary classical variables} and are to have unambiguous quantum
analogs.

ii) Associate with each $f$ in ${\cal S}$ an {\it elementary quantum
operator} $\hat{f}$ and consider the free associative algebra
generated by these abstract operators.  Impose on this algebra the
(generalized) canonical commutation relations
\be [\hat{f}, \hat{g}] = i \hbar \widehat{\{ f,g\} }\ee
for all $f$ and $g$ in ${\cal S}$. In addition, if the set ${\cal S}$
is over-complete, impose on the algebra also `anti-commutation
relations', namely the relations that capture the algebraic relations
that exist between elements of ${\cal S}$. For instance if $f$, $g$
and $h=fg$ are all in ${\cal S}$, then $\hat{f}\cdot\hat{g} +
\hat{g}\cdot\hat{f}= 2 \hat{h}$.  Denote the resulting associative
algebra by ${\cal A}$.

iii) Introduce an involution, $*$, on ${\cal A}$ by setting
\be (\hat{f})^{*}=\hat{\bar{f}}\ee
for all elementary variables $f$ (the bar denotes complex conjugation
as before) and requiring that $*$ satisfies the defining properties of
an involution: $(\hat{A} + \lambda \hat{B})^{*}=\hat{A}^{*}+
\bar{\lambda}\hat{B}^{*}$; $(\hat{A}\hat{B})^{*}= \hat{B}^{*} 
\hat{A}^{*}$ and $(\hat{A}^{*})^{*}=\hat{A}$, for all $\hat{A}$, 
$\hat{B}$ in ${\cal A}$ and complex numbers $\lambda$. Denote the
resulting $*$-algebra by ${\cal A}^{*}$.

iv) Choose a linear representation of ${\cal A}$ on a complex vector
space $V$. (The $*$-relations are ignored at this step).

v) Introduce on $V$ an inner product $\l\, ,\,\r$ so that the ``quantum
reality conditions'' are satisfied
\be \l\Psi, \hat{A}\Phi \r= \l\hat{A}^{*}\Psi, \Phi\r\label{qrc}\ee
for all $\Phi$, $\Psi$ in $V$ and $\hat{A}$ in ${\cal A}^{*}$. Thus,
it is the $*$-relations that are to select the inner product.

The program requires two external inputs: the choice of ${\cal S}$ in
step (i) and the choice of the carrier space $V$ of the representation
in step (iv). If the choices are viable, i.e. if the program can be
completed at all, the resulting inner product is unique on each
irreducible sector of the representation of ${\cal A}$ on $V$
\cite{ran}. In the framework of this program, the textbook treatments 
of field theories correspond to choosing for elements of ${\cal A}$
the smeared field operators, and, for $V$, the Fock space or,
alternatively, suitable functionals of fields. In the loop
quantization, on the other hand, one changes this strategy. both ${\cal S}$
and $V$ are now constructed from holonomic loops.
\goodbreak

\subsection{Algebra based on loop variables}
\label{s3.2}
Let us now implement this program for the Maxwell field using loop
variables. Let us define the {\it smeared holonomy} of self-dual
connections as:
\ba
h_r[\alpha] & := & \exp \left({\T\frac{1}{e}} \int \d^3x 
\oint_{\alpha} \d s\, \dot{\alpha}^a(s)\,{}^\uparrow\!\!A_a(x)
f_r(x-\alpha)\right)
\nonumber\\
&=&\exp\left({\T\frac{1}{e}}\int \d^3x\, F^a_r (\alpha, x)\,
{}^\uparrow\!\!A_a(x)\right)
\ea
or equivalently,
\be h_r[\alpha] = \exp\left[-{\T\frac{1}{e}}\int\frac{\d^3k}{|k|} 
(z_1(k)\bar{F}_1(k) + z_2(k)\bar{F}_2(k))\exp\left(-\frac{r^2k^2}{2}
\right)\right]\, .\ee
Being a function of the self dual connection it can be regarded as a
``configuration variable''. As a momentum variable we will take the
(real) electric field $E^a(x)$, or its Fourier transform $E^a(k)$.
(Strictly speaking we should take the smeared observable
$E[f]=\int_{\Sigma} E^a f_a d^3x$, but this smearing will not be
relevant for our results.)  Hence, $h_r[\alpha]$ and $E^a(k)$ provide
us with a (over-) complete coordinatization of the phase space. The
space ${\cal S}$ of elementary classical variables required in the
first step of the quantization program shall be the vector space
generated by the $h_r[\alpha]$ and $E^a(k)$. It is closed under
Poisson-bracket operation because
\be \left\{ h_r[\alpha], E^a(k)\right\} = {\textstyle \frac{i}{e}} 
F^a_r(\alpha, k) h_r[\alpha]\label{loopal}\ee

The next step in the quantization program is the construction of the
algebra ${\cal A}$ of quantum operators. Let us associate with each
$h_r[\alpha]$ in ${\cal S}$ an operator $\hat{h}_r[\alpha]$ and with
each $E^a(k)$ an operator $\hat{E}^a(k)$ and consider the associative
algebra generated by finite sums of products of these elementary
quantum operators. On this algebra impose the commutation relations:
\ba\left[ \hat{h}_r [\alpha], \hat{h}_r [\beta]\right] =  
0 \qquad &;& \qquad \left[ \hat{E}^a (k), \hat{E}^a (k^{\prime}) 
\right] =  0 \nl
\left[ \hat{h}_r [\alpha], \hat{E}^a (k)\right] &=&  
-  {\textstyle \frac{\hbar}{e}} F^a_r (\alpha , k) \hat{h}_r [\alpha]\ea
Furthermore, we must incorporate in this quantum algebra the fact that
$h_r[\alpha]$ is over-complete. i.e. there are algebraic relations
among them; $h_r[\alpha] h_r[\beta]=h_r[\alpha \# \beta]$. This is
achieved by imposing on the algebra the relations $\hat{h}_r[\alpha]
\hat{h}_r[\beta]=\hat{h}_r[\alpha \# \beta]$  for all holonomic loops
$\alpha$ and $\beta$. The result is the algebra ${\cal A}$ of quantum
operators.

\goodbreak
\subsection{Loop representation}
\label{s3.3}
The next step in the program is to choose a vector space $V$ and a
representation of the quantum operators. The procedure involved is
generally exploratory. Thus, one does not specify all the required
regularity conditions right in the beginning; the precise definition
of spaces considered becomes clear only at the end of the
construction. This will also be the case in our construction. 

We wish to choose for $V$ a vector space of suitable functionals of
loops.  As noted at the end of section \ref{s2.2}, in the standard
electric field representation, one can choose states as suitably
regular functionals $\Phi[F]$ of smooth, vector fields $F^a(k)$ which
are transverse, i.e., satisfy $F^a(k) k_a = 0$.  Now, in section
\ref{s2.2}, (for each choice of a smearing function $f_r$) we set up a
mapping $\alpha\mapsto F^a_r(\alpha, k)$ from loops to smooth
transverse vector fields in the momentum space. We can just pull back
the functionals $\Phi(F)$ via this map to obtain certain functionals
$\Psi(\alpha)$ on the loop space:
\be \Psi(\alpha) = \Phi[F]|_{F=F_r(\alpha,k)}\, . \label{rep} \ee
(Using the regularity conditions on $\Phi$ that come from the standard
electric-field representation, it is not difficult to check that the
map has no kernel, i.e., $\Psi(\alpha) = 0 $ for all $\alpha$ if and
only if $\Phi[F] =0$.) Since the transverse vector fields $F^a(k)$
have only two components $F^{\pm}(k)$, from now on we will regard
$\Phi$ as functionals of the two fields $F^{\pm}$.

Thus, for the representation space $V$, we will use the functionals
$\Psi$ on the loop space of the form (\ref{rep}). Using the procedure
that was successful in the loop representation adapted to the
positive-frequency connections \cite{ar2}, the action of the basic
operators $\hat{h}_r[\alpha]$ and $\hat{E}^a(k)$ will be taken to be:
\ba \hat{h}_r[\alpha] \Psi (\gamma) & = & \Psi (\gamma \cdot \alpha)
\nonumber\\
\hat{E}^a(k) \Psi (\gamma) & = & {\T\frac{\hbar}{e}} F^a(\gamma, k) 
\Psi (\gamma)\label{op1}\ea
As is usual in the loop representation, the electric field is diagonal
in the representation. The only non-vanishing commutator between the
basic operators is
\be \left[ \hat{h}_r[\alpha], \hat{E}^a(k)\right] = 
-{\T\frac{\hbar}{e}} F^a_r(\alpha, k)\hat{h}_r[\alpha]\ee
Finally, for later convenience, we note the action of the magnetic
field operators $\hat{B}^\pm$ on these states.
\be  \hat{B}^{\pm}(k) \Psi (\alpha) =\pm e|k|\left. 
\left[\frac{\delta}{\delta F^{\pm}(-k)}\Phi[F^{\pm}(k)]
\right]\right|_{F^{\pm}(k)=F^{\pm}_r(\alpha, k)}\label{op2}\, ,\ee
which is nothing but the ``loop derivative'' evaluated at $F^{\pm}_r$
(see, e.g. \cite{gp}).

Our next task is to find an inner-product on $V$ so that the ``quantum
reality conditions'' (\ref{qrc}) are satisfied. 
Let us begin with an  inner product of the form
\be
\l\Psi | \Psi ^\prime \r := \int \prod_{k, \pm} \d F^{\pm}(k)\,
\e^{-T[F^{\pm}(k)]} \overline{\Phi[F^{\pm}]}\,\Phi^\prime[F^{\pm}]
\label{innpro}
\ee
and determine the measure by imposing the reality conditions. The
property (\ref{formreal}) of form factors implies that
$T[F]$ should be real. It also implies that the reality condition on
the electric field is automatically satisfied. The other condition one
should impose, namely the quantum version of (\ref{mgreal}) is
\ba \langle\Psi|(\hat{B}^{\pm}(k))^\dagger\chi\rangle &=& 
\overline{\langle\chi|\hat{B}^\pm(k)\Psi\rangle}\nonumber\\
&=& \langle\Psi|-\hat{B}^\pm(-k)+2\hat{p}^\pm(-k)|\chi\rangle\, .\ea
Using the form of the operators (\ref{op1}) and (\ref{op2}) for
$\hat{p}^{\pm}(k)$ and $\hat{B}^{\pm}(k)$, we conclude that the
reality condition (\ref{mgreal}) is satisfied if and only if
\be \frac{\delta T}{\delta F^\pm(k)}=\mp\frac{2\hbar}{e^2|k|}
\overline{F^\pm(k)}\, . \ee
The solution to this equation is: 
\be T[F] = -\frac{2\hbar}{e^2} \int \frac{\d^3k}{|k|}
\left[|F^+_r(k)|^2-|F^-_r(k)|^2\right] \ee
Hence, the explicit form of the inner product (\ref{innpro}) is given
by:
\be \langle\Psi|\Psi^\prime \rangle=\int \prod_{k, \pm} \d F^{\pm}(k) 
\,\e^{\left[ \frac{2\hbar}{e^2}
\int \frac{\d^3k}{|k|}\left(|F^+(k)|^2-|F^-(k)|^2\right)\right]}
\overline{\Phi[F^{\pm}]}\,\Phi^\prime[F^{\pm}]\label{innprod} 
\label{ip}\ee
Notice that the basic form of (\ref{ip}) is the same as that of the
inner-product for a free-field in the configuration (i.e.,
Schr\"odinger) representation.
\footnote{Although $F^\pm (k)$ are complex-valued, they arise as
Fourier components of a {\it real} field $F^a(x)$ and hence satisfy
the reality conditions $\bar{F^\pm(k)} = - F(-k)$. The configuration
space underlying our loop representation is thus real and states
$\Phi(F^\pm)$ are {\it arbitrary} complex-valued functions of $F^\pm$
(i.e., not subject o any ``holomorphicity'' condition.)}
There are, however, two important differences. First, our states are
functionals of loops rather than of a configuration field variable
(such as the connection or the electric field).  Second, for the
positive component, the Gaussian is exponentially growing rather than
damping. Hence, while we can take the states to be polynomials in
$F^-$ as in the Schr\"odinger representation, we have to assume that
they are exponentially damped in their dependence on $F^+$.  Thus, for
example, we can take elements of $V$ to be the functionals
$\psi(\alpha)$ on the loop space of the form:
\be \Psi(\alpha) =  P[F^{\pm}_r(\alpha, k)] \exp\left[-\frac{\hbar}
{e^2} \int \frac{\d^3k}{|k|}\left(|F^+_r(k)|^2\right)\right]
\label{states}\ee
where $P[F^{\pm}_r(\alpha, k)]$ is a polynomial in $F^{\pm}$. As
usual, the Cauchy completion will enlarge this space; the Hilbert
space of all states will contain more general functionals.  In this
description, $F^+$ captures positive helicity while $F^-$ captures the
negative helicity of the photon. Thus, as one might have expected from
our use of only the self-dual part of the connection, the description
is asymmetric in the two helicities.

To summarize, the elementary operators are $\hat{h}_r(\alpha)$ and
$\hat{E}^a$. The space of quantum states is given by functionals
$\Psi(\alpha)$ of holonomic loops which are normalizable with respect
to the inner-product (\ref{ip}) and the action of the elementary
operators is given by (\ref{op1}). For every $r>0$, this loop
representation is naturally isomorphic to the Fock representation%
\footnote{If we let $r$ go to zero, the smearing function $f_r(x)$ 
tends to the $\delta$-distribution and the thickend loop $\alpha_r$
reduces to the loop $\alpha$. However, now the exponent $T$ in the
measure diverges and the loop representation ceases to exist.}
(where the isomorphism, however, depends on the value of $r$.)  The
fact that we are using a loop representation adapted to self-dual
connections is reflected in the measure that dictates the inner
product (\ref{ip}).  In the loop representation adapted to
positive-frequency fields \cite{ar2}, for example, the measure has the
same form but the squares of both $|F^\pm|$ appear with negative signs
in the exponent.
\goodbreak

\section{Measure and the Gauss linking number}
\label{s4}

Recall that our quantum states are functionals of thickened loops
$\alpha_r$, or equivalently, of their form factors $F^\pm_r (\alpha,
k)$; it is for technical convenience that in the intermediate stages
of calculations that we extended them to functionals on the vector
space of all fields $F^\pm(k)$. Therefore, it is instructive to
examine the measure that dictates the inner-product also directly in
terms of the thickened loops. This is easy to achieve: we can just
pull-back the ``Gaussian'' $\exp -T$ that dictates the inner-product
to the space of thickened loops. The result is trivially given by:
\be \exp\left( -T[F_r(\alpha,k)]\right) = \exp \left[\frac{2\hbar}{e^2} \int 
\frac{\d^3k}{|k|} \left(|F^+_r(\alpha, k)|^2-|F^-_r(\alpha, k)|^2
\right)\right] \label{pullback}\ee
We will now show that this loop functional can be expressed in terms
of the Gauss linking number.

Let us begin by recalling the definition of the linking number.  Given
non-intersecting loops $\alpha$ and $\beta$ the Gauss linking number
${\cal GL}(\alpha, \beta)$ between them can be expressed in terms of
their form factors as:
\be {\cal GL} (\alpha,\beta)=\int \d^3x F^a(\alpha, x) w_a(\beta, x)\ee
where $ F^a(\alpha, x)$ is the form factor for $\alpha$ and
$w_a(\beta, x)$ is a potential for the form factor of $\beta$:
$\epsilon^{abc}\partial_b w_c(\beta, x)=F^a(\beta, x)$.  The integral
is independent of the specific choice of the potential $\omega_a
(\beta, x)$ because $F^a(\alpha, x)$ is divergence free.  Note that
neither the definition of the form factor $F^a(\alpha, x)$ nor that of
the potential $\omega_a(\beta, x)$ requires any background fields on
the underlying oriented 3-manifold $R^3$; in particular, there is no
reference to the 3-metric. (Since $F^a$ is a vector density, the
$\epsilon^{abc}$ in the definition of $\omega_a(\beta, x)$ is the
Levi-Civita density which is naturally available on any oriented
3-manifold.) This is to be expected since the Gauss linking number is
a topological invariant.

Nonetheless, one can use the flat metric $q_{ab}$ on $R^3$ to express
the linking number in more familiar terms.  First, we have the
well-known form used by Gauss himself\cite{gauss}:
\be {\cal GL}(\alpha,\beta):=\frac{1}{4\pi}\int \d s\int \d t
\,\epsilon_{abc}
\dot{\alpha}^a(s)\, \dot{\beta}^b(t) \frac{\alpha^c(s)-\beta^c(t)}
{|\alpha(s) - \beta(t)|^3}\label{glno} \ee
For our purposes, a more convenient form is the one involving the
Fourier transforms of the form factors. The Fourier transform of the 
potential has the form:
\ba F^a(\beta, k)&=& iw_c(\beta, k)k_b\epsilon^{abc}\nl
&=& w_c(\beta, k) |k|(m^a\bar{m}^c - \bar{m}^am^c)\nl
&=& |k| (m^a w^+(\beta, k)  -\bar{m}^aw^-(\beta, k))\ea
whence,
\be F^+(\beta, k) = |k|w^+(\beta, k) , \quad {\rm and} \quad 
F^-(\beta, k) = -|k|w^-(\beta, k) .  \ee
Therefore, the Gauss linking number takes the form
\ba {\cal GL}(\alpha, \beta) &=&  \int \d^3k \bar{F}^a(\alpha, k) 
w_a(\beta, k)\nl
&=&  \int \d^3k (\overline{F^+(\alpha, k)m^a} + 
\overline{F^-(\alpha, k)\bar{m}^a}) w_a(\beta, k)\nl
&=&\int \frac{\d^3k}{|k|} (\overline{F^+(\alpha, k)}F^+(\beta, k) -
\overline{F^-(\alpha, k)}F^-(\beta, k)) \ea

Finally, we will need the notion of Gauss number of the
thickened loops $\alpha_r$ and $\beta_r$.  This is just the total
linking number of loops in $\alpha_r$ with those in $\beta_r$:
\ba  {\cal GL} (\alpha_r, \beta_r) &:=& \int \d^3y \int \d^3z\, f_r(y) 
f_r(z)\,{\cal GL}(\alpha^a + y^a, \beta^a + z^a) \nonumber\\
&=&\int \frac{\d^3k}{|k|} (\overline{F^+_r(\alpha, k)}F^+_r(\beta, k) 
- \overline{F^-_r(\alpha, k)}F^-_r(\beta, k)) \ea
Hence the self-linking number of a thickened loop $\alpha_r$ is given 
by:
\be {\cal GL}(\alpha_r, \alpha_r)=
\int \frac{\d^3k}{|k|} \left[\overline{F^+_r(\alpha, k)}F^+_r(\alpha, k) -
\overline{F^-_r(\alpha, k)}F^-_r(\alpha, k)\right]\,  \ee
whence the ``Gaussian'' (\ref{pullback}) on the space of thickened
loops which dictates the inner-product can be expressed as:
\be \exp \left(-{T[F_r(\alpha,k)]}\right) = \exp \,
\left[\frac{2\hbar}{e^2}{\cal GL}(\alpha_r, \alpha_r)\right]\ee
This is the result we were seeking. (Note, incidentally, that the
coefficient of the linking number is $2$ over the fine structure
constant.)

We conclude with a remark.  Had we used positive frequency connections
\cite{ar2}, for example, the  loop functional (\ref{pullback}) would have
been replaced by
\[ \int \frac{\d^3k}{|k|} (\overline{F^+_r(\alpha, k)}F^+_r(\alpha, k) +
\overline{F^-_r(\alpha, k)}F^-_r(\alpha, k))\,  \]
which has no obvious interpretation in terms of the topology of loops.
Similarly, if we had worked in the self-dual connection
representation, the measure would have been dictated by a ``Gaussian''
on the space of {\it connections} (see chapter 11.5, especially Eq 42b
in \cite{aa:91} and \cite{rov-sm}) and would therefore also have had
no relation to topological invariants of loops. We need both
self-duality of the connection {\it and} the loop representation to
relate the photon inner product with the Gauss linking number.

\bigskip
\centerline{\bf Acknowledgments}

This work was supported in part by the NSF grant 95-14240, by the
Eberly research fund of Penn State University and by DGAPA of UNAM.

\end{document}